\documentclass[twocolumn,nobibnotes,
amssymb,aps,superscriptaddress,floatfix,longbibliography]{revtex4-2}
\usepackage[dvips]{graphicx}% Include figure files
\usepackage{dcolumn}% Align table columns on decimal point
\usepackage{bm}% bold math
\usepackage{amsmath}
\usepackage{xcolor}
\usepackage{float}
\graphicspath{{./figs/}}
\usepackage[compact]{titlesec}
\titlespacing{\section}{5pt}{*2}{*2}
\newcommand{\pluss}{$^+ $}

\newcommand{\callfig}{Fig. }

\begin{document}
%\title{Strong Field Molecular Ionization of Water II:  Does it Represent a Vertical Transition?}
%\title{Strong Field Ionization of Water II:\\
%Few-cycle Dynamics and Strong-Field Effects En Route to Double Ionization}

\title{Strong Field Ionization of Water II:\\
Electronic and Nuclear Dynamics En Route to Double Ionization}

\author{Chuan Cheng}
\affiliation{Department of Physics, Stony Brook University, Stony Brook, NY 11794, USA}

\author{Zachary L. Streeter}
\affiliation{Chemical Sciences Division, Lawrence Berkeley National Laboratory, Berkeley, CA 94720, USA}
\affiliation{Department of Chemistry, University of California, Davis, CA 95616, USA}

\author{Andrew J. Howard}
\affiliation{Stanford PULSE Institute, SLAC National Accelerator Laboratory\\
2575 Sand Hill Road, Menlo Park, CA 94025, USA}
\affiliation{Department of Applied Physics, Stanford University, Stanford, CA 94305, USA}

\author{Michael Spanner}
\affiliation{National Research Council of Canada, 100 Sussex Drive, Ottawa K1A 0R6, Canada}
\affiliation{Department of Physics, University of Ottawa, Ottawa K1N 6N5, Canada}

\author{Robert R. Lucchese}
\affiliation{Chemical Sciences Division, Lawrence Berkeley National Laboratory, Berkeley, CA 94720, USA}

\author{C. William McCurdy}
\affiliation{Chemical Sciences Division, Lawrence Berkeley National Laboratory, Berkeley, CA 94720, USA}
\affiliation{Department of Chemistry, University of California, Davis, CA 95616, USA}

\author{Thomas Weinacht}
\affiliation{Department of Physics, Stony Brook University, Stony Brook, NY 11794, USA}

\author{Philip H. Bucksbaum}
\affiliation{Stanford PULSE Institute, SLAC National Accelerator Laboratory\\
2575 Sand Hill Road, Menlo Park, CA 94025, USA}
\affiliation{Department of Applied Physics, Stanford University, Stanford, CA 94305, USA}
\affiliation{Department of Physics, Stanford University, Stanford, CA 94305, USA}

\author{Ruaridh Forbes}
\email{ruforbes@stanford.edu}
\affiliation{Stanford PULSE Institute, SLAC National Accelerator Laboratory\\
2575 Sand Hill Road, Menlo Park, CA 94025, USA}
\affiliation{Department of Physics, Stanford University, Stanford, CA 94305, USA}
\affiliation{Linac Coherent Light Source, SLAC National Accelerator Laboratory, Menlo Park, California 94025, USA}

\begin{abstract}
We investigate the role of nuclear motion and strong-field-induced electronic couplings during the double ionization of deuterated water using momentum-resolved coincidence spectroscopy. By examining the three-body dicationic dissociation channel, D$^{+}$/D$^{+}$/O, for both few- and multi-cycle laser pulses, strong evidence for intra-pulse dynamics is observed. The extracted angle- and energy-resolved double ionization yields are compared to classical trajectory simulations of the dissociation dynamics occurring from different electronic states of the dication. In contrast with measurements of single photon double ionization, pronounced departure from the expectations for vertical ionization is observed, even for pulses as short as 10~fs in duration. We outline numerous mechanisms by which the strong laser field can modify the nuclear wavefunction en-route to final states of the dication where molecular fragmentation occurs. Specifically, we consider the possibility of a coordinate-dependence to the strong-field ionization rate, intermediate nuclear motion in monocation states prior to double ionization, and near-resonant laser-induced dipole couplings in the ion. These results highlight the fact that, for small and light molecules such as D$_2$O, a vertical-transition treatment of the ionization dynamics is not sufficient to reproduce the features seen experimentally in the coincidence double-ionization data.
%[ADDED NOTE FROM ZACHARY:] I think the last sentence should say "intense field coincidence double-ionization data" because the Frank-Condon treatment certainly is sufficient for coincidence double-ionization data using a single photon.
\end{abstract}
\maketitle

%%%%%%%%%%%%%%%%%%%%%%%%%%%%%%%%%%%%%%%%%%%%%%%%%section 1
\section{Introduction}
Investigations into the mechanisms of strong-field ionization (SFI) remain an important cornerstone in ultrafast science research due to its central role in strong-field induced phenomena such as high-harmonic generation \cite{paul2001observation,popmintchev2012bright,marangos2016development}, laser-induced electron diffraction \cite{meckel2008laser,amini2019imaging,blaga2012imaging} and Coulomb-explosion imaging \cite{stapelfeldt1998time,bocharova2011charge}. SFI has also been used as a ``pump" for experiments that aim to study charge transfer or charge migration in molecular cations \cite{kraus2015measurement,kubel2016steering,sabbar2017state,ramasesha2016real}. For molecular systems this continued interest is in part due to the complexity of the SFI process when compared with atomic systems \cite{kjeldsen2005strong,lin2017comparison}. Critical differences that are partially responsible for this are: the spatial arrangement of the nuclei that give rise to the molecular structure; the presence of internal degrees of freedom, vibrations and rotations; and the significantly higher density of electronic states that are typically found in these systems \cite{lezius2001nonadiabatic,kjeldsen2005influence}. 

%[ADDED NOTE FROM PHIL:]   NOTE, HOWEVER, THAT ANDY'S PRESENTATION OF THE DATA IS VIA NEWTON PLOTS, WHICH ARE SUMS OVER ALL POLARIZATION.  SO THIS WAY OF SEPARATING THE TWO PAPERS ISN'T QUITE RIGHT. HERE THERE IS A LOT OF FOCUS ON BETA-KER PLOTS INSTEAD OF NEWTON PLOTS, BUT THESE SHOW ESSENTIALLY THE SAME PHYSICS.  PERHAPS A MORE SIGNIFICANT  DIFFERENCE HERE IS THAT THERE IS A CRITICAL COMPARISON OF THE DATA TO ZACH'S CALCULATIONS, WHICH ARE BASED ON INSTANTANEOUS DICATION FORMATION, WHEREAS ANDY ASSUMES THAT THERE MUST BE SOME INTERNAL MOTION RIGHT FROM FIGURE 1, AND THEN USES THE DATA PLUS CALCULATIONS FROM MICHAEL AND VINOD TO CHARACTERIZE THAT MOTION.

%For many experiments, the ideal scenario is that SFI serves as a sudden ``vertical" transition, in which a portion of the ground state vibrational wave function is  projected onto the electronic states of the mono- or dication. 
For many experiments, the ideal scenario is that SFI serves as a ``sudden'' transition, where the nuclei do not move during the pulse that induces the ionization \cite{holmegaard2010photoelectron,schell2018sequential,sandor2016strong}. The electron rearrangement that accompanies sudden transitions can be compared to the ``vertical'' transitions that describe weak-field perturbative ionization, where a portion of the ground state vibrational wave function is  projected onto the electronic states of the mono- or dication. 
The notion of a vertical transition is based on the idea that the transition dipole moment for the coupling of two electronic states by a weak (perturbative) external field is roughly independent of nuclear coordinate over the extent of the initial wave function. This allows one to factor the coupling matrix element into an electronic term which is multiplied by an overlap of the initial vibrational wave function with the vibrational eigenstates of the upper electronic state, and thus the initial vibrational wave function is simply ``vertically lifted" and projected (mapped) onto the excited state potential energy surface (PES). This assumption is frequently described in terms of the Franck-Condon principle \cite{coolidge1936study}.

%SFI experiments have aimed to achieve vertical transitions by using very short pulses to ionize, for which there is no time for the nuclei to move during the pulse \cite{holmegaard2010photoelectron,schell2018sequential}. However,
Nuclear motion during the pulse is not the only mechanism for non-vertical ionization. Coordinate-dependent strong-field ionization rates \cite{Urbain2004,fang2008strong} and impulsive Raman excitation in one of the electronic states could also result in deviations from vertical ionization \cite{kraus2013high,liu2018near}. 

 %This is different from the idea of a ``Franck-Condon region", which refers to the region of the PES over which the initial wave function extends. In our work, we are examining the first idea rather than the second. 

Recent work that considered the single-photon double ionization of water showed excellent agreement between the measured and calculated momentum resolved yield of fragment ions assuming vertical ionization \cite{streeter2018dissociation,reedy2018dissociation}.  Here we explore the extent to which SFI with short pulses can be considered vertical. Using a combination of coincidence velocity map imaging of fragment ions and trajectory calculations for the three body dissociation dynamics of the molecular dication, we demonstrate that even for very short pulses ($<$ 10~fs), the ionization cannot be considered to be ``vertical", but involves reshaping (changes to the first and second moments of the distribution) of the vibrational wavefunction during the ionization dynamics. We discuss different contributions to the wave function reshaping during ionization.

Our measurements, calculations and analysis may help interpret previous work that made use of SFI as probe of excited state dynamics \cite{horton2018strong,Forbes2017,Ding2019} and for vibrational wave packet holography \cite{petersen2004control}. In the case of probing excited state dynamics, while SFI produced qualitatively similar time dependent yields as weak field or single photon ionization, quantitative agreement was not possible \cite{horton2018strong}. We argue that this is due to variation in SFI rates with nuclear geometry and reshaping of the vibrational wave packet during SFI. While the holography measurements showed very nice agreement between the experiment and theory in terms of the interference fringe positions and visibility, details of the comparison were not perfect due to the limited ion imaging ability and the assumption of vertical ionization. Our work directly examines this assumption and can be thought of as characterizing the instrument response function associated with SFI as a probe. Particularly for early time delays in pump probe experiments, where the initial and final PESs vary significantly with nuclear coordinate, SFI can differ from vertical ionization significantly. Also, for high intensities, one can see enhancement of ionization to higher charge states through charge resonance enhanced ionization, which also violates the notion of vertical ionization \cite{legare2005laser,hong2015charge}. Finally, for pulses longer than 10~fs, experiments and simulations are greatly affected by laser-induced dynamic alignments and couplings between different electronic states \cite{mccracken2020ionization,koh2020ionization,howard2021strong,cheng2020momentum}.
 
In terms of theoretical investigations into the SFI dynamics of water, recent frozen-nuclei time-dependent R-matrix {\it ab initio} ionization computations on H$_2$O \cite{RMatrixH2O} suggest that laser coupling in the ion can significantly modify the angular dependence of the ionization dynamics when using multi-cycle pulses. However, if intermediate motion takes place on the ionic surface, adding the laser coupling during the ionizing pulse without also including nuclear motion on the laser-coupled ionic states becomes suspect. Motion on the ionic states will affect the phases and populations on each laser-coupled surface, which can in turn affect the ionization dynamics. The current alternative approach to modelling ionization in multi-cycle strong fields when motion in the ion states is present can be seen in studies of N$_2^+$ lasing, where the strategy is to compute half-cycle ionization yields which are then used as inputs into separate laser-coupled ion dynamics computation \cite{richter2020,lytova2020}. This is the approach that we adopt in the present manuscript---our half-cycle ionization yields were presented in a previous paper \cite{cheng2020momentum}, and herein we carry out the laser-driven ionic dynamics. A more rigorous treatment could involve adding the ionization contributions from different half-cycles to the laser-coupled ion dynamics simulation through inclusion of a source term in a density matrix approach \cite{zhang2020}. 
%{\it (Adding some discussion to this effect makes the importance of the intermediate motion more relevant from a theory perspective maybe?) (Maybe this whole paragraph could be put in summary/conclusions?)}

\begin{table*}[]
\begin{tabular}{|c|c|c|c|c|c|}
\hline
C$_s$ sym & spin state order & C$_{2\nu}$ sym & C$_{2\nu}$ config & 2-body branching ratio {[}\%{]} & 3-body branching ratio {[}\%{]} \\ \hline
X$^3$A''  & T$_{0}$               & $^3$B$_1$     & (3a$_1$1b$_1$)$^{-1}$ & 92                              & 8                               \\ \hline
2$^3$A''  & T$_{1}$               & $^3$A$_2$     & (1b$_2$1b$_1$)$^{-1}$ & 0                               & 100                             \\ \hline
1$^3$A'   & T$_{2}$               & $^3$B$_2$     & (1b$_2$3a$_1$)$^{-1}$ & 0                               & 100                             \\ \hline
1$^1$A'   & S$_{0}$               & 1$^1$A$_1$     & (1b$_1$)$^{-2}$    & 100                             & 0                               \\ \hline
1$^1$A''  & S$_{1}$               & $^1$B$_1$     & (3a$_1$1b$_1$)$^{-1}$ & 87                              & 13                              \\ \hline
2$^1$A'   & S$_{2}$               & 2$^1$A$_1$     & (3a$_1$)$^{-2}$    & 26                              & 74                              \\ \hline
2$^1$A''  & S$_{3}$               & $^1$A$_2$     & (1b$_2$1b$_1$)$^{-1}$ & 0                               & 100                             \\ \hline
3$^1$A'   & S$_{4}$               & $^1$B$_2$     & (1b$_2$3a$_1$)$^{-1}$ & 0                               & 100                             \\ \hline
3$^1$A''  & S$_{5}$               & 3$^1$A$_1$     & (1b$_2$)$^{-2}$    & 0                               & 100                             \\ \hline
\end{tabular}
\caption{\textcolor{red}{} Table of electronic states of the water dication in different conventions. Branching ratios are for the dissociation dynamics on the \textit{ab initio} potential surfaces with initial conditions from the Wigner phase space distribution of the ground vibrational state,  calculated from $10^5$ trajectories on each potential surface, similar to the calculations in  \cite{reedy2018dissociation,streeter2018dissociation}.}
\label{table:state_convention}
\end{table*}

%%%%%%%%%%%%%%%%%%%%%%%%%%%%%%%%%%%%%%%%%%%%%%%%%section 2
\section{Experiment}\label{sec:experimental_setup}
Two similar apparatuses were used to carry out the measurements. Since they are almost the same, we provide a description for the apparatus used to carry out the 10~fs pulse measurements (described in detail in previous work \cite{zhao2017coincidence,cheng2019electron}) and indicate any differences for the apparatus used for the 40~fs pulse measurements. Briefly, the output from a commercial amplified Ti:sapphire laser system (1~mJ, 780~nm, 1~kHz) is spectrally broadened using filamentation in Ar gas, and compressed to $\sim$10~fs using chirped mirrors and an acousto-optic pulse shaper \cite{dugan1997high}. The laser pulses are directed into a vacuum chamber (base pressure of $10^{-10}$~mbar) and focused at the center of a velocity-map imaging (VMI) spectrometer using an in-vacuum concave silver mirror ($f$~=~5~cm). The \textit{in situ} intensity was calibrated by SFI of argon, measuring the classical 2U$_{p}$ cutoff for field-ionized electrons from argon \cite{corkum1993plasma} at low pulse energies and extrapolating to higher ones (using a procedure outlined in Ref.~\cite{bryan2006atomic} for the 40 fs measurements). The estimated intensities for the 10~fs and 40~fs measurements are 400~TW/cm$^2$ and 600~TW/cm$^2$, respectively.

Target D$_{2}$O molecules are expanded into a separate source chamber and subsequently skimmed to yield an effusive molecular beam \cite{zhao2017coincidence}. This beam intersects the focused laser at the center of the electrostatic lens stack of the VMI spectrometer \cite{zhao2017coincidence}. The extracted ions and electrons are recorded using a microchannel plate (MCP), phosphor screen, and TPX3CAM camera \cite{zhao2017coincidence,Nomerotski2019} with 1.5 ns time resolution (a Roentdek hexanode detector, with a time resolution of $<$1~ns \cite{jagutzki_multiple_2002} was used for the 40~fs pulse measurements). 

In order to ensure the low count rates required for coincident detection of all charged particles, a working pressure of 4$\times$10$^{-10}$~Torr was used throughout the experiment (with a base pressure of about 1$\times$10$^{-10}$~Torr). The adopted pressure resulted in an average event rate of approximately 0.8 per laser shot, which in turn corresponded to an event rate of $<$ 0.1 per shot for the double-ionization channels considered in this work. Presented in Table~\ref{table:statistics} are the experimentally extracted yields and branching ratios for the dissociation channels following strong-field double ionization. Two-body fragmentation into OD$^{+}$/D$^{+}$ is the dominant channel followed by three-body fragmentations into D$^{+}$/D$^{+}$/O and D$^{+}$/D/O$^{+}$. In the present work we focus our discussion on the three body D$^{+}$/D$^{+}$/O fragmentation channel. A detailed analysis of fragmentation into OD$^{+}$/D$^{+}$ was included in an earlier publication, which investigated the roles of dynamic and geometric alignment during water double ionization \cite{howard2021strong}. We note that our extracted ratios differ significantly in some channels from reported values for single photon double ionization \cite{pedersen2013photolysis} and, more critically, SFI with higher intensity longer duration pulses \cite{zhao2019strong}. Furthermore, we note the absence of the weak D$_{2}^{+}$/O$^{+}$ channel observed in Ref.~\cite{zhao2019strong} for our data recorded with 10~fs pulses. A detailed investigation into the pulse duration dependence of double ionization branching ratios is beyond the scope of the present work. 

\begin{table}[]
\begin{tabular}{|c|c|c|c|c|}
\hline
channels & \begin{tabular}[c]{@{}c@{}}counts: \\ uncorrected\end{tabular} & \begin{tabular}[c]{@{}c@{}}efficiency: \\ uncorrected\end{tabular} & \begin{tabular}[c]{@{}c@{}}counts: \\ corrected\end{tabular} & \begin{tabular}[c]{@{}c@{}}efficiency: \\ corrected\end{tabular} \\ \hline
D+/OD+   & 6.5x10$^4$    & 0.632                                                              & 3.7x10$^5$                                                      & 0.702                                                            \\ \hline
D+/D+/O  & 3.0x10$^4$                                                        & 0.288                                                              & 8.8x10$^4$                                                      & 0.165                                                            \\ \hline
D+/O+/D  & 8.2x10$^3$                                                        & 0.080                                                              & 7.0x10$^4$                                                      & 0.133                                                            \\ \hline
\end{tabular}
\caption{Table of branching ratio of relevant water dication dissociation channels. The detection efficiency of different ions are estimated to be 0.58(D$^+$), 0.2(O$^+$) and 0.3(OD$^+$).}
\label{table:statistics}
\end{table}

\section{Trajectory Calculations\label{sec:D2O_DD_trajectory_sim}}
%Revisions in progress by CWM -3-24-2021
Classical trajectory calculations simulate the conditions of single-photon double ionization of the ground state of D$_2$O in which the molecule undergoes a Franck-Condon transition to the doubly ionized excited state. The trajectories are propagated on the potential surfaces of the lowest nine states of the water dication. Those states are created, in the simple molecular orbital picture, by removing two electrons in all possible ways from the highest three filled molecular orbitals of the neutral as indicated in Table \ref{table:state_convention}, leaving the D$_2$O$^{++}$ ion in all possible spin states for each configuration. The surfaces were calculated earlier~\cite{gervais2009h,streeter2018dissociation} with MOLPRO~\cite{werner2012WIRE,MOLPRO} using internally contracted multireference configuration interaction (icMRCI) methods at the configuration interaction singles and doubles (CISD) level, including the Davidson correction to the CI energy. The full dimensional surfaces were then fitted using a functional form developed by Gervais \textit{et al.} \cite{gervais2009h}.

\begin{figure}
	\centering
	\includegraphics[width=\linewidth]{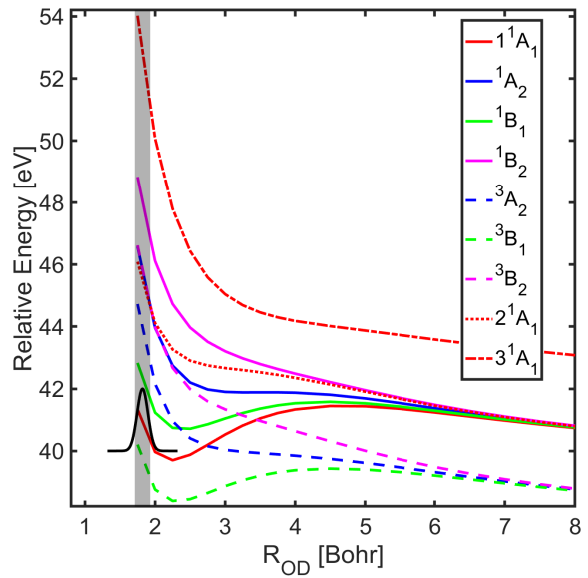}
	\caption{Potential energy curves for the first nine states of the D$_{2}$O di-cation as a function of the symmetric stretch coordinate $R_{\textrm{OD}}$, together with the neutral ground state wave function (solid black line). The vertical shaded grey line illustrates the extent of the ground state wave function. The curves included in this figure are reproduced from Ref. \cite{streeter2018dissociation}.}
	\label{fig:PES}
\end{figure}

The trajectory calculations assumed the neutral water molecule is initially in its ground vibrational state with zero total rotational angular momentum. The initial conditions for $10^5$ trajectories on each potential surface were sampled from the Wigner phase space distribution for the computed normal modes of the ground electronic state in icMRCI calculations.  The Phase space distribution has the form 
\begin{equation}\label{WignerDistribution}
    W(\mathbf{Q},\mathbf{P}) = \frac{1}{ \left(\pi \hbar  \right)^{3N-6} } \prod_{j=1}^{3N-6} e^{-\frac{\omega_j}{\hbar} Q_j^2 -\frac{1}{\hbar \omega_j}P_j^2} 
\end{equation}
where $N$ is the number of atoms, $w_j$ are the associated frequencies, and the vectors $\mathbf{Q}$ and $\mathbf{P}$ are the normal mode coordinates and momenta, respectively, and is positive definite for the case of the ground state.  A similar trajectory study was performed for double ionization of  H$_2$O previously~\cite{streeter2018dissociation} and compared extensively with experimental momentum images~\cite{reedy2018dissociation} from single photon double ionization measurements using the cold target recoil ion momentum spectroscopy (COLTRIMS) technique~\cite{ullrich1997JPB,dorner2000PR,ullrich2003RPP}.
That comparison validated the accuracy of this \textit{ab initio} treatment of the dissociation of the cation in all but its finest details.  The present calculations differ only in the masses of the atoms (D versus H) and the larger number of trajectories on each surface ($10^5$ versus $10^3$).  They thus show small quantitative differences from the previous work, for example in the two-body versus three-body dissociation branching ratios.

The potential energy curves along the symmetric stretch coordinate of the first 9 states of dication are shown in the Fig.~\ref{fig:PES}. The state characters are labelled according to C$_{2\nu}$ symmetry in that figure. Outlined in Table~\ref{table:state_convention} are the equivalent labelling conventions in other symmetries, together with the dominant electronic configurations for these states near the equilibrium geometry of neutral molecule and the associated two- and three-body theoretical branching ratios for the present case of D$_2$O. The criterion for categorizing a trajectory as three body in this work was that the R$_{\mathrm{OD}}$ distance of one deuterium be $200$~Bohr or greater and that the other reach at least $50$~Bohr. All others were categorized as two-body dissociation.

\section{Measurements}

%Recent fragment ion momentum resolved measurements of the one photon double ionization of water were beautifully captured by classical calculations of the dynamics on the first 9 states of the dication \cite{streeter2018dissociation,reedy2018dissociation}, as seen by comparing the calculated and measured coincidence H\pluss/H\pluss/O yield as a function of angle between the two H$^+$ (D$^+$ in this work) ions' momentum, $\beta$, and the kinetic energy release (KER). 
%\textcolor{red}{[CWM question: In my understanding $\beta$ is  the angle between the final momenta of the H$^+$ ions and not the H-O-H angle, no? Also we should probably use H$^+$/H$^+$/O here because that study was for water.]} % The state resolved double ionization channels (D\pluss/D\pluss/O) of deuterated water can be seen from an earlier work\cite{streeter2018dissociation,reedy2018dissociation} with the help of weak field single photon double ionization.
%By using the energy conservation and having access to the kinetic energy of the two electrons, this earlier work was able to separate contributions from different oxygen atom electronic states in the dissociation product, which is directly linked to the initial ionization state. Moreover, the D\pluss/D\pluss/O coincidence plotted as a function of $\beta$ and KER ($\beta$-KER plot) exhibits distinctive state resolvable features, as can be seen in the Appendix Fig.~\ref{fig:Appd_stateIslands}.

\begin{figure}
	\centering
	\includegraphics[width=\linewidth]{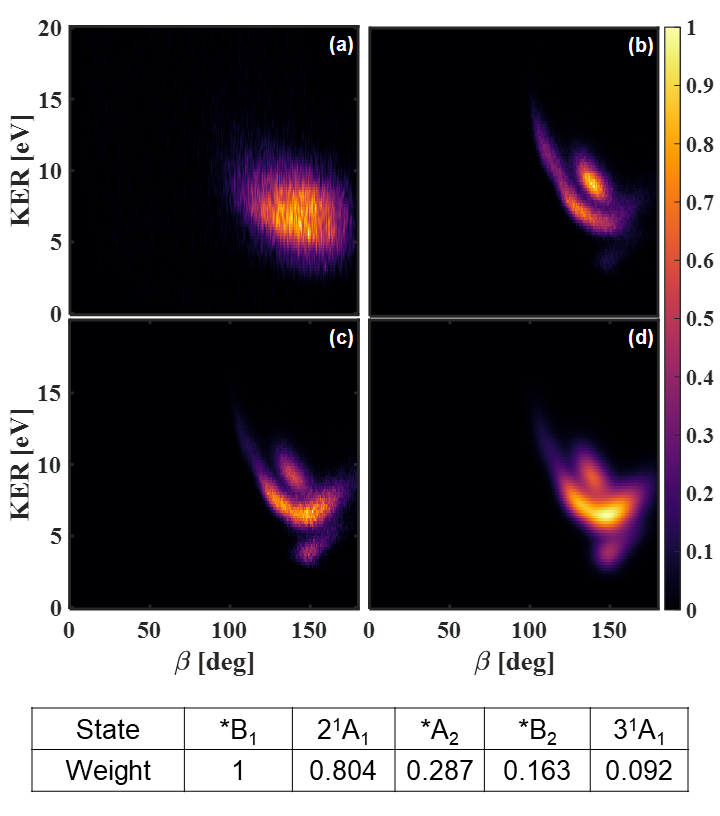}
	\caption{Measured and calculated D$^+$/D$^+$/O yield as a function of angle between the two D$^+$ ions' momentum, $\beta$, and the kinetic energy release (KER). Panel (a) Measured coincidence D$^+$/D$^+$/O yield as a function of $\beta$ and KER for laser parameters described in the text; (b) Simulated D$^+$/D$^+$/O yield as a function of $\beta$ and KER for equal population of each of the first nine states of the dication; (c) simulation $\beta$-KER with fit; (d) simulation $\beta$-KER with experimental resolution included. The table at the bottom lists the relative fitting populations for each states where the states are labeled with their C$_{2\nu}$ symmetry due to lack of the spin information in the observable. Details are outlined in the text.}
	\label{fig:beta_KER_w_weighting}
\end{figure}

Given the agreement between the trajectory calculations and the single photon double ionization measurements described above, the question we address here is whether strong-field double ionization prepares a similar superposition of states of the dication, and whether the wave packet launched on each state via SFI is similar to the one launched by single photon (weak field) ionization - i.e. can the SFI process be thought of as vertical? We therefore compare the calculated double ionization yield as a function of $\beta$ and KER with our measurements. Fig.~\ref{fig:beta_KER_w_weighting} shows the measured and calculated D\pluss/D\pluss/O yield. Panel (a) shows the measured yield as a function of $\beta$ and KER. Panel (b) shows the calculated yield vs $\beta$ and KER assuming vertical ionization with equal population of the di-cation states. Panel (c) shows the same results as panel (b) with coefficients for the first 9 states of the dication fitted to achieve the best agreement with the measurements. Panel (d) shows the calculated results accounting for experimental broadening of the features due to the limited resolution of our VMI apparatus. Details of which are discussed below.

In the fitting procedure, due to the lack of the spin state information about the oxygen atom, states of the the same C$_{2\nu}$ symmetry (e.g. $^1$B$_1$ and $^3$B$_1$) have been grouped together since they are not separable in the $\beta$-KER plot \cite{streeter2018dissociation}. From now on we just use the word ``state" to represent different C$_{2\nu}$ symmetries. As can be seen from \callfig~\ref{fig:Appd_stateIslands}, each state populates a distinct region (``island") of the $\beta$-KER plot. More details on the island assignments can be found in Refs. \cite{reedy2018dissociation,streeter2018dissociation} as well as in Appendix~\ref{Appd_stateIslands}. Based on these islands, a simple principle component analysis (PCA) procedure that minimizes the residual was used to reproduce the $\beta$-KER plot. Different states population have been fitted and the reconstructed $\beta$-KER plot is shown as \callfig \ref{fig:beta_KER_w_weighting}(c). The fitted relative state population has been listed as the table in the figure as well. The coefficients from the fit are decreasing with increasing ionization potential, roughly in agreement with expectations from a simple quasi-static tunnel ionization model of Ammosov-Delone-Krainov (ADK theory \cite{tong2002theory}). An interesting observation is that while the relative weights for states that involve removing electrons from different orbitals are in reasonable agreement with predictions from ADK theory, the relative weights for singlet states that involve removing two electrons from the same orbital with opposite spins are significantly higher than predictions based on ADK theory. This observation may indicate cooperative behavior, or non-sequential double ionization to states of the dication that involve removing two electrons from the same orbital.

The fitting in \callfig \ref{fig:beta_KER_w_weighting}(c) only coarsely matches the measurements shown in \callfig \ref{fig:beta_KER_w_weighting}(a). Although the experimental yield shows a center of mass similar to the calculations, the distinct state resolved features in the $\beta$-KER plot are not captured in the SFI results. An important first check that we carried out in addressing this discrepancy is to determine whether our experimental resolution broadens the features such that they are no longer resolved. Using the coincidence events from 2 body dissociation channel (D\pluss/OD\pluss), we obtained an uncertainty for the measured D$^+$ momentum of 2.7~a.u by checking their nature of momentum conservation in all three dimensions (p$_x$, p$_y$ and p$_z$). The energy of each D$^+$ ion in the 3 body channel is about 4.5~eV. Thus the uncertainty of KER is calculated to be $\delta \textrm{KER} = \sqrt{2}p\delta p/m = 0.99$~eV. Similarly one can obtain the uncertainty of angle between the two D$^+$ momentum to be $\delta \beta = \sqrt{2}\delta p/p = 6.24^{\circ}$.

Applying the PCA, together with the experimental resolution correction, yields Fig.~\ref{fig:beta_KER_w_weighting}(d). The agreement between theory and experiment is still relatively poor: Notably, the islands corresponding to the $^1$B$_1$ and $^3$B$_1$ states are absent in the measurements (see Appendix \ref{Appd_stateIslands}). In the panel Fig.~\ref{fig:beta_KER_w_weighting}(d), where the states are blurred according to our instrument response function, the state islands feature are still present. This is completely at odds with the experimental yield shown in Fig.~\ref{fig:beta_KER_w_weighting}(a), where no discrete features are present. Thus, we conclude that there has to be some mechanism that drives the difference between simulation and the experiment. We note that this disagreement between experiment and theory is independent of exactly what representation one chooses (i.e. which observables to look at), and a number of different data representations are shown in the Appendix Sec.~\ref{SecAppend:alternatives}.%We also tried some alternative ways to compare the experiments and the simulation. None of them really give good agreements. Yet there are some interesting aspects in those plots so we decided to put them in the appendix. From now on we need to think about the reason behind the differences.\par

\begin{figure}[ht]
	\centering
	\includegraphics[width=1.05\linewidth]{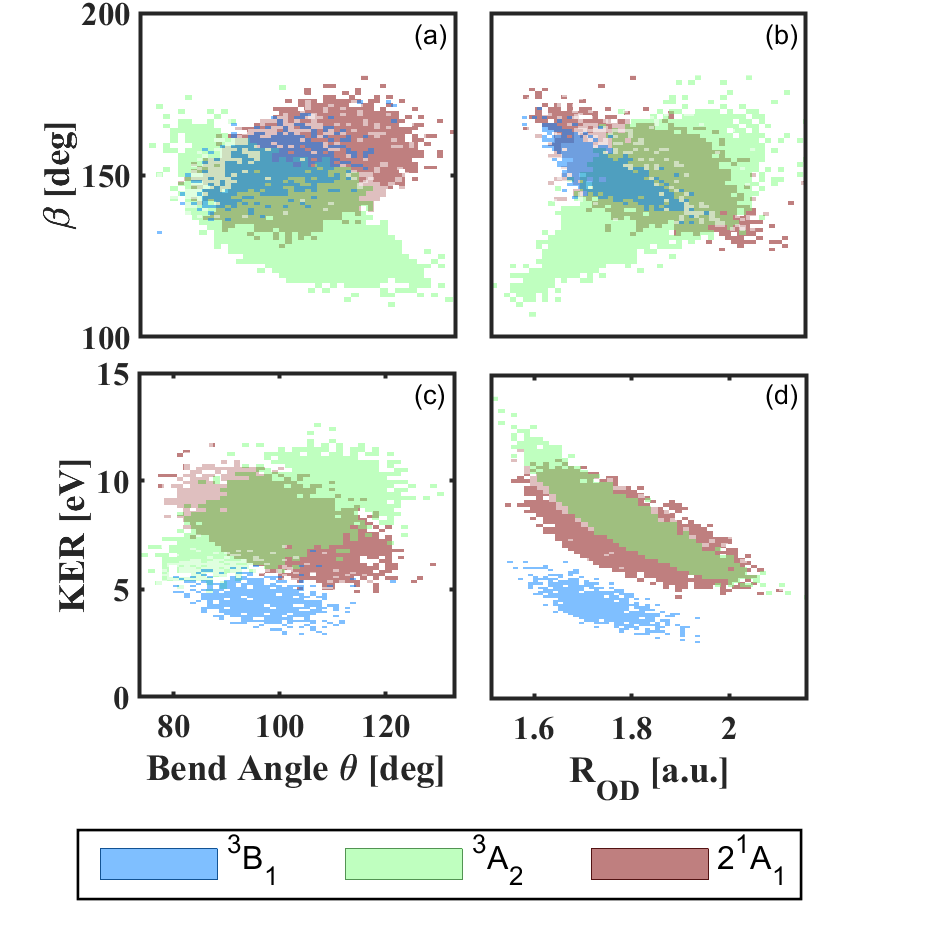}
		\caption{Calculated D\pluss/D\pluss/O yield as a function of KER, $\beta$ and initial O-D separation or D-O-D bend angle $\theta$ for three states. The top left panel (a) shows the D\pluss/D\pluss/O yield as a function of initial D-O-D angle and final $\beta$. Panel (b) shows the D\pluss/D\pluss/O yield as a function of initial O-D distance and final $\beta$. Panel (c) shows the D\pluss/D\pluss/O yield as a function of initial D-O-D angle and final KER. Panel (d) shows the D\pluss/D\pluss/O yield as a function of initial O-D distance and final KER. The color coding here is consistent with \callfig\ref{fig:Appd_stateIslands}.
		}
	\label{fig:ini_vs_fin}
\end{figure}

%\subsection{How does the final Beta and KER correlate with different initial Beta values or other internal coordinates in the N dimensional Wigner distribution}

\section{Dynamics Calculations and Discussion}\label{Sec: CalcAndDiss}
The calculated and measured $\beta$-KER plots show some rough agreement, but disagree on many details, so it is natural to ask whether the discrepancies can be due to dynamics occurring during the SFI process. In particular, we consider the role of wave-packet reshaping (due to coordinate-dependent ionization rates for example) and nuclear dynamics (such as bending or rotation). As a first test of how much displacement the wave packet would need to alter the $\beta$-KER plot, we mapped the correlation between initial and final values for the distribution of sample points used in the classical trajectory calculations. As the dynamics calculations include an ensemble of initial and final coordinate values, we can plot the final values as a function of the initial ones, allowing us to see if there are correlations between them that lead the strong field reshaping of the ground state wave function to smear out the features in the measured $\beta$-KER plot. %Both the initial and final values are already extracted from the single photon dissociation trajectory simulations. Thus by plotting the correlations between initial reaction coordinates and final observables, we could see if there are correlations between them that could smear out the features in the measured $\beta$-KER plot as a result of strong field reshaping of the ground state wave function. 

Based on the simulation we constructed the correlation maps shown in \callfig\ref{fig:ini_vs_fin}. Presented are correlation maps for three different electronic states, illustrating the correlation between initial opening angle and symmetric stretch coordinate with respect to the final $\beta$ and KER. The correlation between initial symmetric stretch coordinate R$_{\textrm{OD}}$ and final $\beta$ and KER, or initial DOD angle and final $\beta$ and KER are a result of the finite width of the initial wave function and the nonzero slope of the dicationic PES at the Franck Condon location. %As one can see from the right two panels, the final $\beta$ angle and KER have strong dependence on initial symmetric stretch coordinate.

%[Zach Comment:] The paragraph below says ^3A_2 breaks down axial recoil approximation. The ^3B_1 and 2 ^1A_1 break down axial recoil so this needs to be revised.
The upper row in Fig. \ref{fig:ini_vs_fin}, which shows how the final $\beta$ values depend on the initial coordinates, exhibits a strong state dependence. The $^3$B$_1$ and $2^1$A$_1$ states show a positive correlation between the final $\beta$ value and the initial opening angle, while the $^3$A$_2$ state shows a negative correlation.  The behavior with respect to the initial O-D distance is the opposite.  This illustrates how reshaping or motion of the initial ground state wavefunction can result in different $\beta$ distributions than those predicted for single photon ionization. The bottom row in \callfig\ref{fig:ini_vs_fin} shows the KER dependence on the initial coordinates. Again, there  is a significant dependence of the KER on initial bend angle and O-D distance, illustrating how reshaping or motion of the initial ground state wavefunction can distort the KER distributions predicted for single photon ionization. These plots motivate an examination of the different strong-field mechanisms that can result in non-vertical ionization and the measured $\beta$-KER plots. The sensitivity of the correlation between initial and final coordinates to electronic state is related to the breakdown of the axial recoil approximation and ``slingshot" motion of the D$^+$ ions for the $^3$B$_1$ and $2^1$A$_1$ states, as discussed in more detail in  \cite{streeter2018dissociation}.

%illustrating ``slingshot" motion of the two D$^+$ during dissociation, in which the two D$^+$ ions come out with negative momentum along the direction bisecting the two D atoms in the neutral molecule.  Their potential energy curves are illustrated in Fig. 5 of Ref \cite{streeter2018dissociation}. This type of motion represents a dramatic breakdown of axial recoil approximation which is discussed in great detail in Ref \cite{streeter2018dissociation,reedy2018dissociation,gervais2009h}. The other states exhibiting negative correlation between final $\beta$ and initial opening angle, like $^3$A$_2$, fulfill the axial recoil approximation. The bottom row in \callfig\ref{fig:ini_vs_fin} shows the KER dependence on the initial coordinates, illustrating how reshaping or motion of the initial ground state wavefunction can result in different KER distributions than those predicted for single photon ionization. This plot motivates an examination of the different strong-field mechanisms that can result in non-vertical ionization and the measured $\beta$-KER plots.

%%%%%%%%%%%%% Start of Michael's results %%%%%%%%%%%%%%%%%%%%

The calculated $\beta$-KER plot shown in Fig.~\ref{fig:beta_KER_w_weighting} (b) and Fig.~\ref{fig:Appd_stateIslands} are results of applying one-photon perturbation theory to describe the transition from the initial neutral directly into the dication states, which is then followed by classical propagation to compute the final fragment energies and angles. In this treatment, the initial Wigner function launched on the dication states is then simply the Wigner function corresponding to the normal-mode ground state of the neutral species, that is, the Franck-Condon wave packet. While this treatment is applicable to weak-field ionization where one-photon perturbation theory is applicable,  additional effects are present during SFI that modifies the initial neutral ground-state wavepacket before it arrives on the dication surfaces. First, the SFI rate can depend strongly on the nuclear coordinates, an effect which can reshape the initial ground state nuclear wavepacket during ionization. Second, multiple ionization via SFI is typically a sequential process where there is a time delay between the ejection of each liberated electron, thereby giving the nuclei a chance to relax and rearrange in between different ionization events. Finally, since there is still a strong-field present during this time delay, near-resonant laser-driven electronic transitions can occur, causing additional non-ionizing electronic transitions that reshape the nuclear wave packets while the molecule is in an intermediate ionic state.  

We now outline each of these effects in more detail. We do not attempt a complete treatment of D$_2$O double ionization in strong-fields with all degrees of freedom active, which although desirable, represents a massive theoretical and computational task. Rather, we limit ourselves to outlining each effect using simplified reduced-dimensionality models. Since ionization to both the X and A cation states is expected to occur \cite{cheng2020momentum}, modifications and dynamics arising from ionization to both X and A are used to exemplify these effects.

\begin{figure}
\centering
\includegraphics[width=0.95\columnwidth]{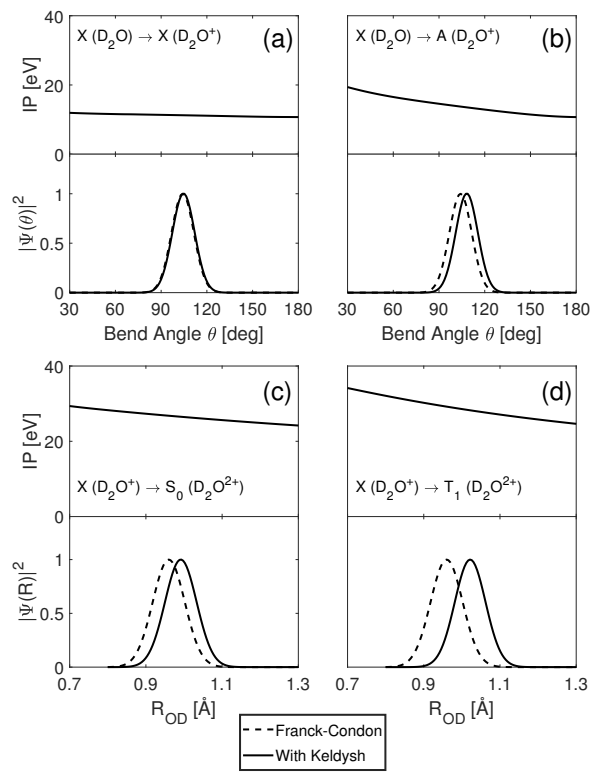}
\caption{Modification of the nuclear wave packet due to the 
    coordinate-dependent Keldysh ionization rate. Panels (a) and
    (b) show the effect of the Keldysh weighing on the bending
   coordinate during the single ionization event to both the X and A
cation states. The panels plot both the ionization potentials (IP) for the transitions in question, the unmodified initial neutral
	(i.e. Franck-Condon) wave packet, and the nuclear wave packet
	after applying the Keldysh weighting. Panels (c) and (d) show
	the effects of the Keldysh weighting for two examples 
	 of single-to-double ionizing transitions (indicated in the 
	 figures) as a function of $\mathrm R_{\mathrm{OH}}$ bond length of the symmetric stretch mode. All probability densities $|\Psi|^2$ have been normalized have a maximum of 1.}
    \label{fig:keldysh}
\end{figure}

{\it Coordinate-dependence of the SFI rate}:  
SFI with low-frequency fields can be envisioned as a quasistatic tunnel ionization process where the ionization rate depends exponentially on the ionization potential (IP) through the Keldysh tunneling rate $\Gamma({\mathbf R},t)$ \cite{Keldysh,PPT}.
In this description, ionization occurs in short sub-cycle bursts near the peaks of the laser oscillations. With the assumption that the nuclei remain stationary during a single ionization burst, the ionized wave packet after the burst can be written as
\begin{equation}\label{EqKeldyshWeighting}
    \Psi_{K}({\mathbf R}) = \Gamma({\mathbf R,t}) \Psi_{0}({\mathbf R}).
\end{equation}
where $\Psi_{0}({\mathbf R})$ is the initial nuclear wave packet before ionization and $\mathbf R$ stands for all nuclear coordinates. If ionized from the neutral at equilibrium, $\Psi_{0}({\mathbf R})$ is often called the Franck-Condon wavepacket. The Keldysh rate is given by
\begin{equation} \label{EqKeldyshRate}
    \Gamma({\mathbf R},t) = {\cal P} \exp\left[ -\frac{2}{3} \frac{(2\;{\mathrm{ IP}}({\mathbf R}))^{3/2}}{ |F(t)| }  \right] 
\end{equation}
where ${\cal P}$ is a slowly-varying (i.e. non-exponential) prefactor that depends weakly on IP, $F$, and ${\mathbf R}$. Here $F$ denotes the amplitude of the electric field. In molecular systems, the ${\cal P}$ prefactor would also encode the molecular orientation dependence of SFI as well as other molecular effect such as enhanced ionization, and accurate computation of ${\cal P}$ would require some form of {\it ab initio} numerical simulation of the ionization process.  In the following we set ${\cal P}=1$ for simplicity and consider only the effects of the Keldysh exponent. Eq.(\ref{EqKeldyshWeighting}) shows that the Keldysh rate can modify the spatial structure of Franck-Condon wave packet through the coordinate-dependence of the IP.  
    
Fig.~\ref{fig:keldysh} shows examples of the Keldysh rate modifying the initial Franck-Condon wave packet in D$_2$O. Fig.~\ref{fig:keldysh}a shows the IPs for the first ionization step from the neutral to the X state of the ion along the bend coordinate $\theta$ together with cuts through the nuclear wave function both with and without applying the Keldysh weighting applied.  Fig.~\ref{fig:keldysh}b plots the same but for the neutral to A state.  While very little change in the wave function occurs for X ionization along this coordinate, it can be seen that the Keldysh weighting has the effect of shifting the Franck-Condon wave packet along the bend coordinate. Panels c and d plot similar cuts for two transitions of the second ionization steps, X$\rightarrow$ 1$^1$A$_1$ (S$_0$) and X$\rightarrow$ $^3$A$_2$ (T$_1$), but now taken along the symmetric stretch coordinate. Due to the increased magnitude and steep coordinate dependence of the IPs for these transitions, the Keldysh-induced shifts of the wave function are more pronounced compared to the previous single-ionization examples.
    
{\it Few-cycle nuclear motion in the ion}: 
Following the first ionization event, the nuclear wave packet can begin to move on the cationic surfaces before the second ionization occurs.  This intermediate motion can change the nuclear wave packets before being projected onto the dicationic states. From our investigations of the cationic surfaces, the dominant motion is expected to be along the bend coordinate. In reality the motion in the cation occurs in the presence of the laser field, but we first consider the effects of field-free motion, which alone can already cause pronounced reshaping of the wave packets. Laser-driven motion in the cation is considered below.

The bending wavepacket dynamics on the X$^2$B$_1$ and A$^2$A$_1$ cationic surfaces is simulated using the following simplified model of D$_2$O$^+$. First, the bond lengths are fixed at the neutral equilibrium values, $R_{eq}$, throughout the dynamics.
Second, the bending is restricted to a single plane of motion, where the overall rotational motion about the center-of-mass is not considered during these wavepacket simulation. Finally, the mass of the oxygen atom is assumed to be infinite, which significantly simplifies the corresponding kinetic energy operator. With these restrictions, the Hamiltonian of the model system is written (in atomic units) as
\begin{equation}
    \label{EqHamil}
    \widehat H(\theta,t) = \left[ {\begin{array}{cc}
                        -\frac{1}{2\mu R_{eq}^2} \frac{\partial^2}{\partial \theta^2} + V_X(\theta) & 0 \\
                        0 & -\frac{1}{2\mu R_{eq}^2} \frac{\partial^2}{\partial \theta^2} + V_A(\theta) 
        \end{array}} \right],
\end{equation}
where $\mu = m_D/2$ is the reduced mass of the bend coordinate, $m_D$ is the mass of atomic deuterium, 
%R_{eq}$ is the equilibrium bond length of the neutral, 
while $V_X(\theta)$ and $V_A(\theta)$ are the potential energy
surfaces of the X$^2$B$_1$ and A$^2$A$_1$ states.  This Hamiltonian is used to solve the time-dependent Schr\"odinger
equation (TDSE) $i \partial_t \Psi(\theta,t) = \widehat H(\theta,t)
\Psi(\theta,t)$.  In order to investigate the effect of wave-packet motion as separate from the Keldysh effects discussed above, we populate both the X and A surfaces during the ionization event starting from the unchanged Franck-Condon wave packet, and hence the initial condition for $\Psi(\theta,t)$ at the moment of
ionization is taken to be the bending ground state, $\psi_0(\theta)$, on the
neutral surface. The TDSE is solved using the Fourier-split-operator technique \cite{Feit82}.

\begin{figure}
    \centering
    \includegraphics[width=\columnwidth]{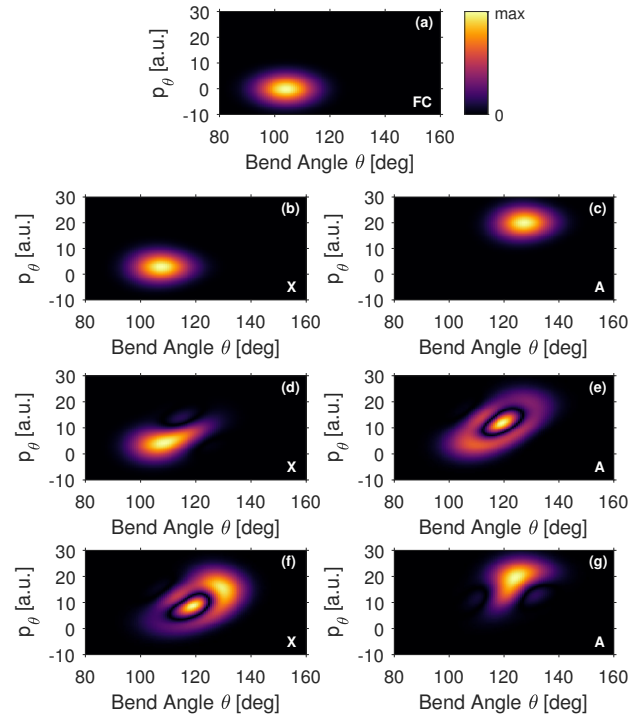}
    \caption{\textcolor{red}{}Wigner functions demonstrating the effects of nuclear dynamics in the cation.
    (a) Wigner function of the Franck-Condon wave packet. (b) \& (c) Wigner functions after field-free propagation on the
    X and A states respectively for a time equal to 2 cycles of the laser.  (d) \& (e) Wigner functions on the X and A states
    after laser-driven propagation for a time of 2 laser cycles with the initial population starting on the X state. The simulation was started at the peak of the pulse ($t=0$) with the laser parameter $\lambda_0$=780~nm, I$_0$ = 400~TW/cm$^2$, and $\tau$=10~fs. (f) \& (g) Same as previous two panels but now with initial population on the A state. Note that in all cases the magnitude of the Wigner function is shown.}
    \label{fig:Wigner}
\end{figure}

The first three panels of Fig.~\ref{fig:Wigner} show the effects of intermediate field-free few-cycle motion on the X and A cation surfaces.  Fig.~\ref{fig:Wigner}a shows the Wigner function of the initial Franck-Condon wave packet.  This is the initial state of the bending coordinate that is used in the $\beta$-KER trajectory simulations. Figs.~\ref{fig:Wigner}b and c plot Wigner functions after this initial state has propagated field-free for a time corresponding to 2 laser cycles of the 780~nm field (2$\times$2.6~fs) on the X and A states, respectively. The X-state Wigner has undergone a little acceleration and motion as can be seen by the small shift of the center of the Wigner function, but there is still significant overlap between the propagated and initial Wigner functions in this case. However, the 2-cycle field-free propagation on the A-state substantially modifies the initial Wigner function, which now has effectively zero overlap with the initial state. The remaining panels of Fig.~\ref{fig:Wigner} pertain to laser-driven motion in the ion, which is outlined in the following.

% It might be informative to add more labels to the panels in Fig.5, if there's room. 
% Something like...

% For (a): "FC, t=0 fs, Laser Off"

% For (b): "X, t=2.6 fs, Laser Off"
% For (c): "A, t=2.6 fs, Laser Off"

% For (d): "X --> X, t=2.6 fs, Laser On"
% For (e): "X --> A, t=2.6 fs, Laser On"

% For (f): "A --> X, t=2.6 fs, Laser On"
% For (g): "A --> A, t=2.6 fs, Laser On"

% ...assuming I've interpreted the caption correctly!

\begin{figure}
    \centering
    \includegraphics[width=7.5cm, trim={2.5cm 0.5cm 2.5cm 0.5cm}]{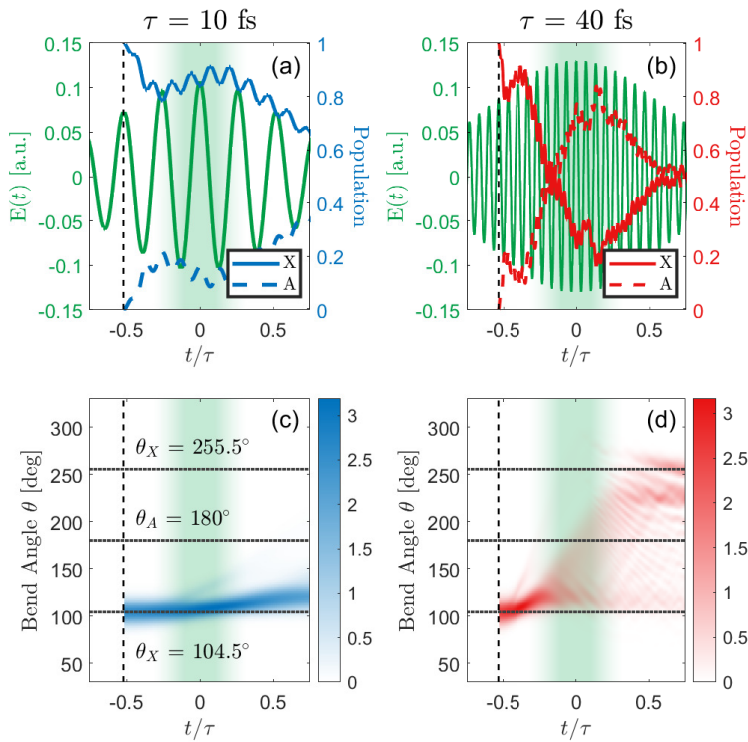}
    \caption{Time-resolved simulations of the strong-field coupling between the X state of the water cation and the A state in the presence of (a) the 10~fs pulse with $\lambda_0$~=~780~nm and I$_0$~=~400~TW/cm$^2$ and (b) the 40~fs pulse with central wavelength $\lambda_0$~=~800~nm and peak intensity I$_0$~=~600~TW/cm$^2$ (parameters chosen to match the experiments). In each case, the cation is initiated in the X state (red/blue solid line) at a local peak in electric field (solid green line) that roughly corresponds to the point at which the pulse intensity is half it's maximum value: $t/\tau$~$\approx$~-0.5. Due to the presence of the field, the X population couples to the A state (red/blue dashed line). Plotted below each of these figures is the probability distribution over bend angle, $\theta$, for the mixture of states displayed above, shown separately for (c) the case of a 10~fs pulse (blue) and (d) the case of a 40~fs pulse (red). In each case, dotted gray lines denote the equilibrium bend-angles for the X and A states. In all four panels, green shading roughly indicates the window in time over which the second ionization event in sequential double-ionization is expected to occur.}
    \label{fig:Coupling}
\end{figure}

{\it Near-resonant dipole coupling in the ion}: 
At 780~nm, the laser induces a near-resonance one-photon coupling between the X and A states. This coupling is included in the wave packet simulation by adding off-diagonal dipole terms to the Hamiltonian, which then becomes
\begin{equation}
    \label{EqHamilF}
    \widehat H(\theta,t) = \left[ {\begin{array}{cc}
                        -\frac{1}{2\mu R_{eq}^2} \frac{\partial^2}{\partial \theta^2} + V_X(\theta) & -F(t)\cdot d_{XA}(\theta)\\
                        -F(t) \cdot d_{XA}(\theta) & -\frac{1}{2\mu R_{eq}^2} \frac{\partial^2}{\partial \theta^2} + V_A(\theta) 
        \end{array}} \right],
\end{equation}
where $d_{XA}(\theta)$ is the transition dipole between these states, and $F(t)$ is the electric field of the laser which is chosen to be parallel to $d_{XA}(\theta)$, i.e perpendicular to the molecular plane.
The electric field of the laser is taken to have a Gaussian envelope
\begin{equation}
    F(t) = {\cal F}_0 \exp \left[ -4 \ln 2 \left( \frac{t}{\tau \sqrt{2}} \right)^2\right] \cos(\omega_0 t)
\end{equation}
where $\omega_0$ is the carrier frequency, ${\cal F}_0$ is the peak electric
field magnitude, and $\tau$ is the full width at half maximum of the
corresponding intensity envelope $|F(t)|^2$.  

The effects of the dipole coupling on the Wigner functions can be seen in Figs.~\ref{fig:Wigner}(d)-(g), while a more complete picture of the laser-driven dynamics is shown in Fig.~\ref{fig:Coupling}. We first consider the Wigner functions. Figs.~\ref{fig:Wigner}(d) and (e) plot the Wigner functions on the X and A states respectively after the system was initialized with the Franck-Condon wave packet on the X state at the peak ($t$=0) of a laser pulse with $\tau$ = 10~fs and intensity of 400~TW/cm$^2$, and allowed to propagate for 2 cycles of the laser. Initially only the X state is populated, but due to the near-resonant dipole coupling some population is transferred to the A state, which is discussed further below. In addition to the modifications due to field-free propagation, seen in Figs.\ref{fig:Wigner})b) and (c), both of the Wigner functions on the X and A states have acquired additional structures after the laser-driven propagation. These structures arise due to light-induced potentials created by the strong near-resonant laser field that modify the field-free potential energy surfaces, an effect known as bond softening \cite{Bucksbaum1990,Sanderson1999}. Additionally, the transfer of population from one state back to the other, which occurs through cascaded one-photon transitions between the X and A states induced by the strong near-resonant laser, also contributes to these structures.  Figs.\ref{fig:Wigner}(f) and (g) plot Wigner functions for the same scenario but now with the initial population starting on the A state. Again, it is seen that new structure not present in the Franck-Condon wave packet have developed.

\begin{figure}
	\centering
	\includegraphics[width=1.1\linewidth]{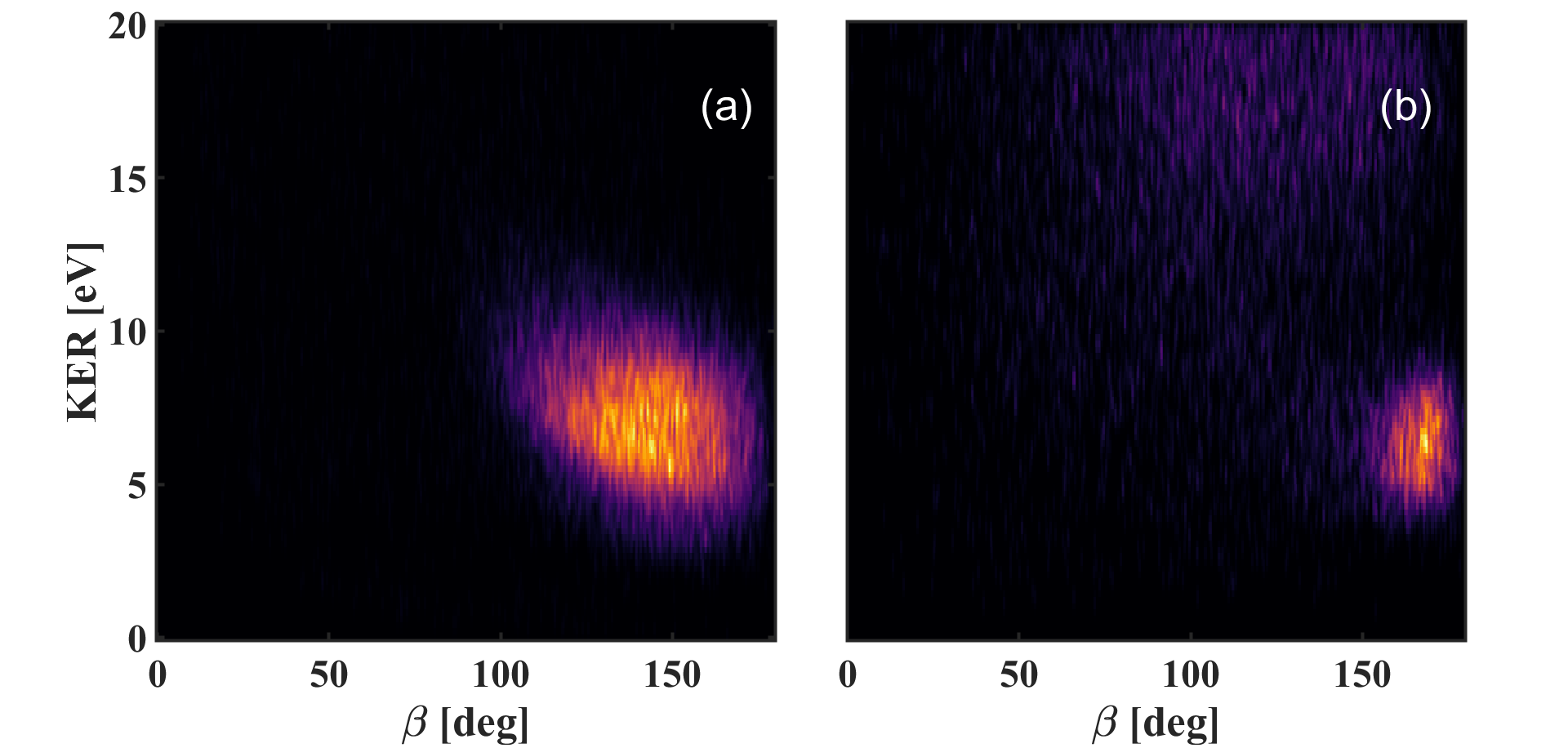}
	\caption{$\beta$-KER plot for SFI with (a) 10~fs and (b) 40~fs pulse durations. A higher kinetic energy feature is observed for the longer pulses in panel (b). This feature can be assigned to trication formation and fragmentation via the D\pluss/D\pluss/O\pluss channel. A discussion of which is presented in Appendix~\ref{sec:trication_formation}.}
	\label{fig:compare40fsVS10fs}
\end{figure}

In order to put things into perspective, we compare the amount of X to A coupling that occurs in a 10~fs pulse to a significantly longer pulse with a duration of 40~fs. Fig~\ref{fig:Coupling} shows a broader picture of the laser-driven dynamics. The top panels show the population of ground and first excited states of the monocation together with the laser field as a function of time for 10~fs (top left) and 40~fs (top right) laser pulses. The bottom two panels show the corresponding probability density as a function of D-O-D angle and time. The figures illustrate the fact that while population transfer and nuclear dynamics can take place during a 10~fs pulse, they have a much more dramatic effect for a 40~fs pulse.  

The calculations described above suggest that in a longer pulse, dynamics in the monocation en route to the dication during a longer pulse also leads to significant unbending and a $\beta$-KER plot that is shifted to larger angles. We therefore compare the $\beta$-KER plot for 10 and 40~fs pulses as a test of the conclusions of the theoretical results. Fig.~\ref{fig:compare40fsVS10fs} shows this comparison, which bears out the predictions based on the calculations above. We note that the comparison shows that 40~fs tend to have larger $\beta$ angles and also the lower KER. The former is what we expected from the simulation while the latter is not easy to explain. Nevertheless, this comparison provides strong evidence that while there is rough agreement between the measured and calculated momentum resolved fragment ion yields for sub 10-fs pulses assuming vertical ionization, reshaping of the wave packet and nuclear dynamics during ionization can lead to significant differences between strong- and weak-field ionization. And one needs to understand the strong-field dynamics in order to predict the reaction products in an SFI measurement. Finally, a weak diffuse feature at higher KER ($\sim$20~eV) and smaller $\beta$ values is observed for 40-fs pulses, which is notably absent for the short pulse case. This feature can readily be assigned to trication Coulomb explosion and is discussed in detail in Appendix~~\ref{sec:trication_formation}.

\begin{figure}
	\centering
	\includegraphics[width=1.1\linewidth]{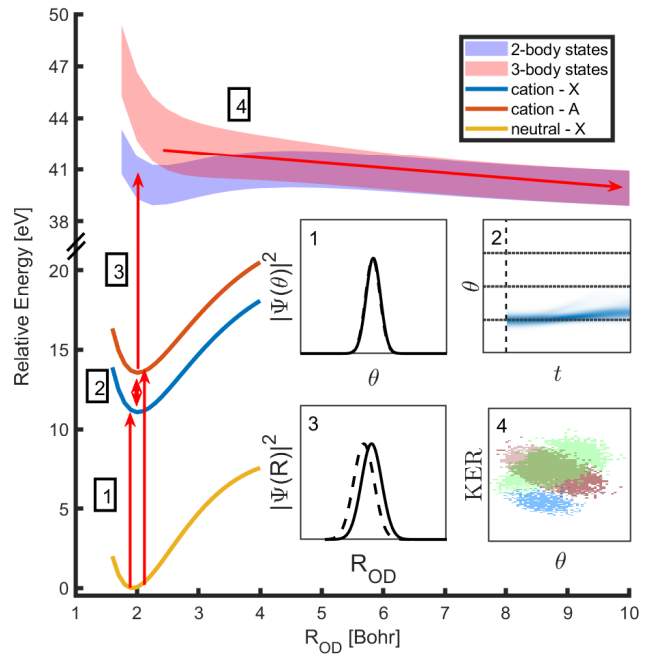}
	\caption{\textcolor{red}{} Concluding figure with PES of relevant states. Four different steps are participating in the strong field double ionization of water: 1- Tunnel ionization to X and A states of monocation, which involves reshaping of wave function due to R-dependent ionization; 2- near resonant coupling of X and A states as well as motion on A state's potential; 3- ionization to the dication, which also involves reshaping of wave function; 4- dication dissociation, which can be simulated through trajectory calculations.}
	\label{fig:PES_concluding}
\end{figure}

\section{Conclusion}
In conclusion, while we find rough agreement between measurements of the D$^+$/D$^+$/O yield as a function of $\beta$ and KER for double ionization of D$_2$O with 10~fs laser pulses and calculations of the same yield assuming weak-field/vertical ionization, there are significant differences in the details. As summarized in the \callfig\ref{fig:PES_concluding}, these differences may be ascribed to wave packet reshaping by the strong-field of the laser pulse, and nuclear dynamics during the pulse. These result in SFI being a non-vertical transition, with an understanding of the strong-field dynamics being important for the calculation of the fragment ion yield momentum distributions.

Our calculations of the wave packet reshaping suggest that the coordinate dependence of the tunnel ionization rate plays an important role in reshaping the initial wave packet, particularly for the second ionization step in a sequential double ionization process. This is a result of the dication having a larger coordinate dependence to the ionization potential. Calculations of the monocation dynamics in the presence of the strong laser field indicate that field dressed nuclear dynamics can also reshape the vibrational wave packet. The reshaped wave packet is then projected onto the dication potential energy surfaces toward the end of the pulse. These observations are supported by measurements of the double ionization yield as a function of $\beta$ and KER for 10~fs and 40~fs pulses. This comparison has significant bearing on the use of SFI as a probe of molecular structure and dynamics, and indicates that while very short pulses ($<$ 10~fs) can minimize nuclear dynamics during the pulse, wave packet reshaping by the strong field of the laser can still result in significant distortion of the initial wave function of the molecule. We believe this work will have important consequences for pump-probe techniques, such as time-resolved Coulomb explosion imaging, that aim to track nuclear dynamics during excited state photochemical processes.

These results showcase the need for a more comprehensive theoretical description of SFI processes, which include the role of nuclear motion occurring during the ionizing laser pulse duration. As outlined in the introduction, such nuclear dynamics could have implications
for computations of SFI from multi-cycle pulses when invoking the frozen-nuclei approximation,  
such as recent state-of-the-art time-dependent R-matrix ionization computations for
H$_{2}$O that highlighted modifications to the angular dependence of ionization when significant laser coupling in the ion is present for the frozen-nuclei case. Further experimental investigations into the pulse shape and intensity dependence of SFI processes in water may help to target future modelling of these dynamics.

\begin{acknowledgments}
We would like to thank Brian M. Kaufman and Yusong Liu for technical support, Spiridoula Matsika and James Cryan for useful discussions. RF, AJH, and PHB were supported by the National Science Foundation. AJH was additionally supported under a Stanford Graduate Fellowship as the 2019 Albion Walter Hewlett Fellow. CC and TW gratefully acknowledge support from the Department of Energy under Award No. DE-FG02-08ER15984.  Work at LBNL was performed under the auspices of the U.S. Department of Energy (DOE), Office of Science, Office of Basic Energy Sciences, Chemical Sciences, Geosciences, and Biosciences Division under Contract No. DEAC02-05CH11231, using the National Energy Research Computing Center (NERSC),  a DOE Office of Science User Facility, and the Lawrencium computational cluster resource provided by LBNL.
\end{acknowledgments}

%%%%%%%%%%%%%%%%%%%%%%%%%%%%%%%%%%%%%%%%%%%%%%%%%section appendix

\appendix
\section{Individual dication state contributions to $\beta$-KER plot}\label{Appd_stateIslands}

\begin{center}
	\begin{figure}
		\centering
		\includegraphics[width=\linewidth]{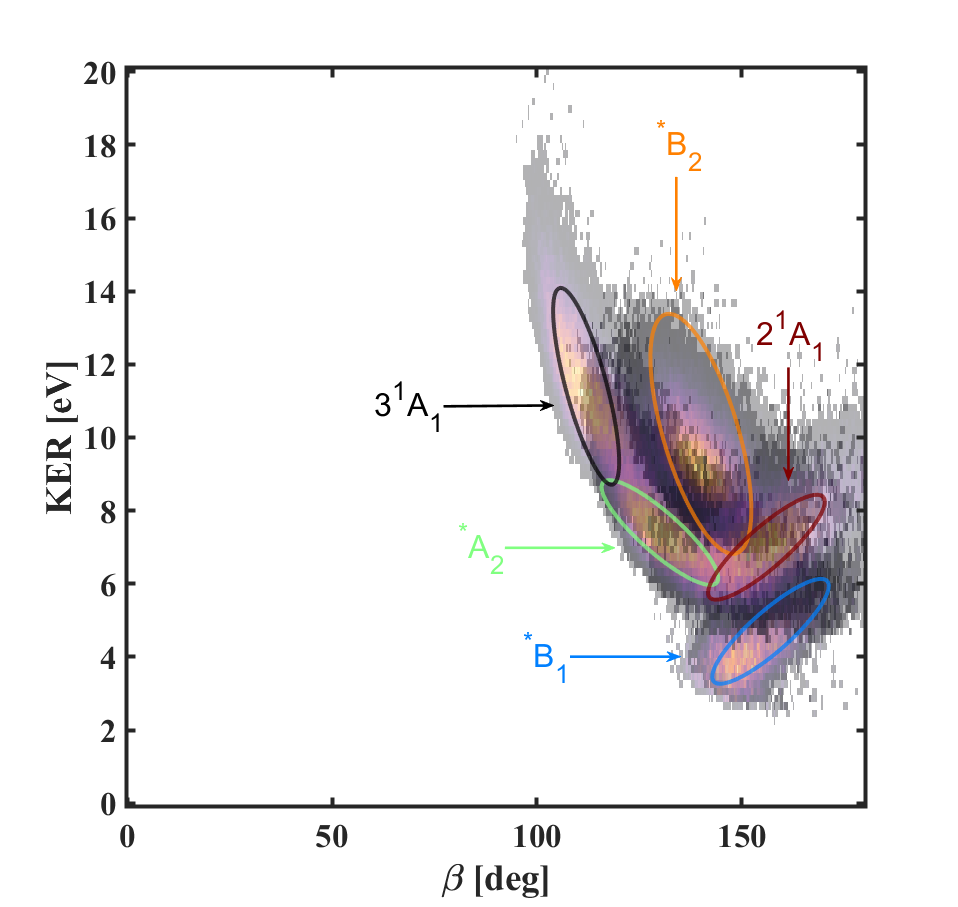}
		\caption{Transparent theoretical $\beta$-KER plot generated from the simulations outlined in Sec.~\ref{sec:D2O_DD_trajectory_sim}. Annotations of the specific states associated with each distinct spectral feature are shown (ovals). A state dependent color coding consistent with Fig.~\ref{fig:ini_vs_fin} is adopted for clarity.}
		\label{fig:Appd_stateIslands}
	\end{figure}
\end{center}

In order to try to connect the observed dissociation features in our strong-field double ionization experiment to specific dication states we adopted a similar methodology to Refs.~\cite{streeter2018dissociation,reedy2018dissociation}. In those studies, the H$^+$/H$^+$/O yield following single photon double ionization of H$_{2}$O were compared against classical trajectory simulations on various electronic states of water diction. By comparing theoretical kinetic energy releases (KERs) and relative angle correlations (between the two H$^{+}$ fragments) to the experimental data, it was unambiguously shown that specific features could be connected to dissociation occurring on different electronic states of the dication. 

Classical trajectory simulations of three-body dissociation of D$_{2}$O$^{2+}$ into D$^{+}$/D$^{+}$/O were performed using a methodology briefly outlined in Sec.~\ref{sec:D2O_DD_trajectory_sim}.
In Fig.~\ref{fig:ini_vs_fin}
theoretical KERs and D$^+$/D$^+$ relative angles, $\beta$, from these calculations are shown alongside the experimental data. In order to connect the features observed in theory to the dication states that are expected undergo three-body decay (see Table~\ref{table:state_convention}), Fig.~\ref{fig:Appd_stateIslands} shows the $\beta$-KER plot with annotated regions corresponding to specific final states. For clarity we adopt the same state specific color scheme outlined in Fig.~\ref{fig:ini_vs_fin}.

% \section{Alternative representations of the timepix3 data}
\section{Alternative representations of the D$^{+}$/D$^{+}$/O channel}\label{SecAppend:alternatives}

\begin{center}
	\begin{figure*}
		\centering
		\includegraphics[width=\textwidth]{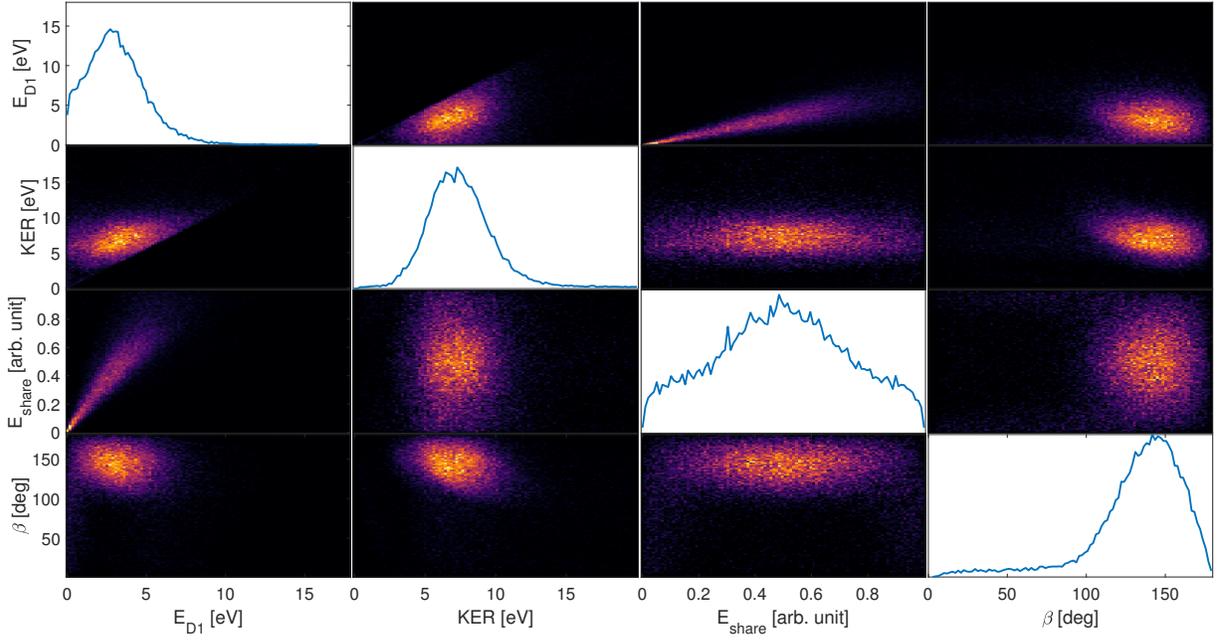}
		\caption{Representations of the multi-dimensional D$^{+}$/D$^{+}$/O coincidence data along various coordinates. Shown are KER, $\beta$, E$_{\textrm{D}1}$ and E$_{\textrm{share}}$ cuts of the data along each of these coordinates in a matrix style plot. Here the diagonal represents the one-dimensional normalized line-out along each of the aforementioned axes.}
		\label{fig:Appd_plotmatrix_exp}
	\end{figure*}
\end{center}

\begin{center}
	\begin{figure*}
		\centering
		\includegraphics[width=\textwidth]{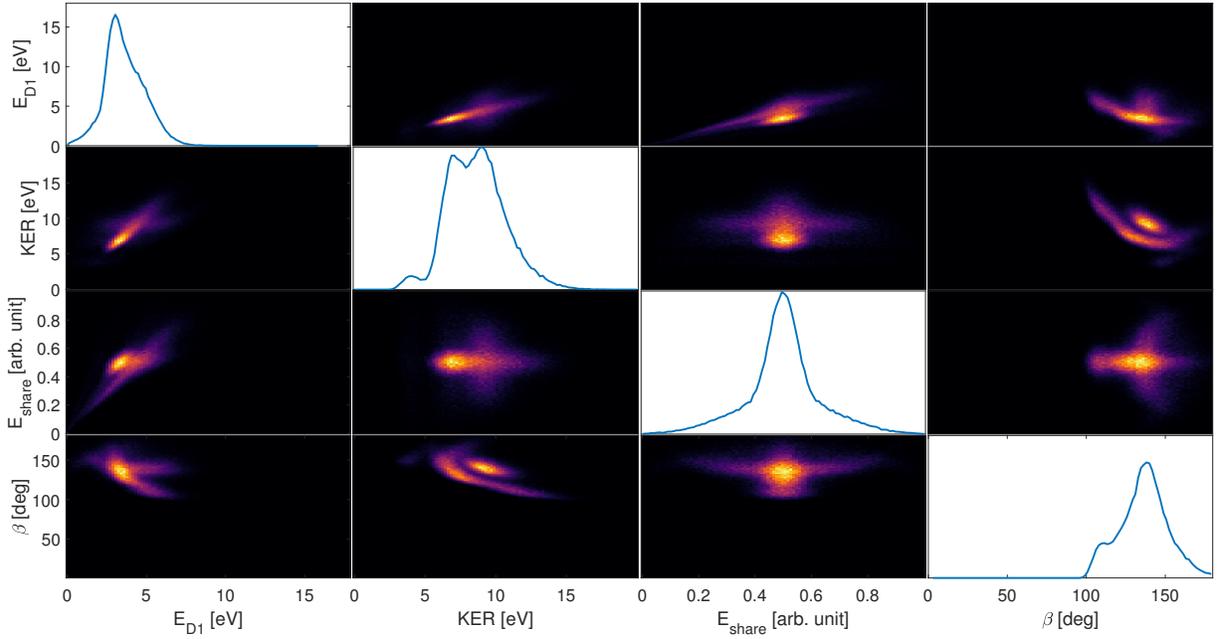}
		\caption{Cuts of theoretical results along the same coordinates outlined in Fig.~\ref{fig:Appd_plotmatrix_exp} for the D$^{+}$/D$^{+}$/O fragmentation channel in the classical trajectory simulations. Details of constraints used to classify the trajectories into specific channels are outlined in Sec.~\ref{sec:D2O_DD_trajectory_sim}.}
		\label{fig:Appd_plotmatrix_simulation}
	\end{figure*}
\end{center}

In Sec.~\ref{sec:experimental_setup} it was briefly outlined that by exploiting the TPX3CAM camera in conjunction with a voltage switching VMI apparatus it is possible to coincidentally detect all charge particles (electrons and ions) following strong-field double ionization, provided a suitably low number ($\ll$1) of double-ionization events per laser shot is achieved. For the ions, the time-stamping capabilities of the camera permit the three-dimensional vector momenta of all fragments to be extracted. This information permits observables, such as the relative angles between fragments or total KERs, to determined. As is outlined in the main text of manuscript, as well as Appendix~\ref{Appd_stateIslands}, a particularly useful way to view the data associated with the D$^{+}$/D$^{+}$/O channel is to consider the relative angle between the two D$^{+}$ ions, $\beta$, as a function of KER. We note, however, that there exists a large number of possible representations of the data due to three-dimensional correlated information extracted during the experiment. To highlight this, Fig.~\ref{fig:Appd_plotmatrix_exp} shows various cuts of the full data set along the KER, $\beta$, E$_{\textrm{D}1}$ and E$_{\textrm{share}}$ coordinates. Here E$_{\textrm{D}1}$ and E$_{\textrm{share}}$ refer to the kinetic energy of the one of deuterons and the sharing of kinetic energy between both the deuterons, respectively. In Fig.~\ref{fig:Appd_plotmatrix_simulation} the equivalent plots for the theoretical simulations outlined in Sec.~\ref{sec:D2O_DD_trajectory_sim} are presented. 

\begin{figure}
	\centering
	\includegraphics[width=\linewidth]{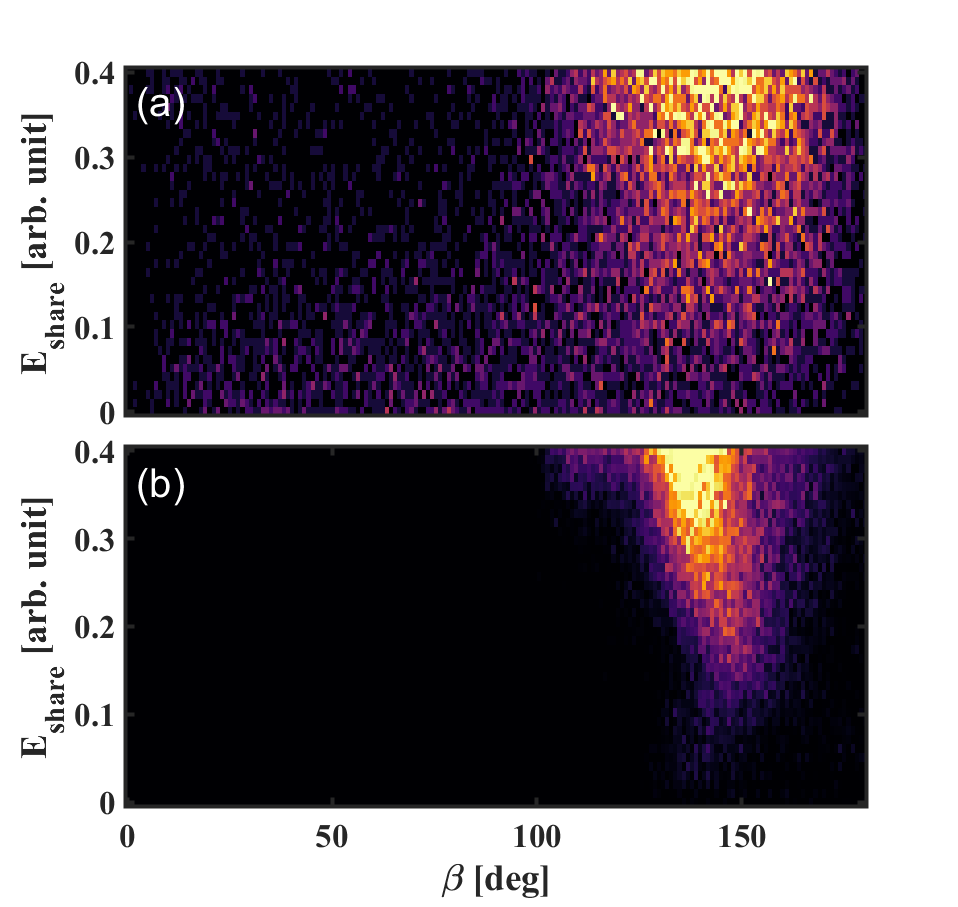}
	\caption{Zoomed version of the $\beta$-E$_{\textrm{share}}$ plots shown in Figs.~\ref{fig:Appd_plotmatrix_exp} and \ref{fig:Appd_plotmatrix_simulation}. A broad feature in angle is observed at E$_{\textrm{share}} < 0.2$ experimentally (a), which in notably absent in the theoretical results (b). This features can be assigned to a sequential dissociation pathway involving predissociation of OD$^{+}$ as discussed in the main text.}
	\label{fig:betaEshare}
\end{figure}

Similar to the discussion of the $\beta$-KER plot in the main text there exists some coarse agreement between theory and experiment in several of the panels. However, it is apparent that the well-resolved features in Fig.~\ref{fig:Appd_plotmatrix_simulation} are absent in the experimental data. This is likely due to the mechanisms outlined in Sec.~\ref{Sec: CalcAndDiss}. A particularly noteworthy region of disagreement is observed in the panel where the yield is plotted as a function of E$_{\textrm{share}}$ and $\beta$. A weak feature extending over all angles at E$_{\textrm{share}}$ = 0 and 1 is seen experimentally but has no analogous signature in the theory. In Fig.~\ref{fig:betaEshare} a zoomed $\beta$-E$_{\textrm{share}}$ plot is presented to highlight this discrepancy. We attribute this large angular spread in $\beta$ to a process by which D$_{2}$O$^{2+}$ initially undergoes two-body decay into D$^{+}$/OD$^{+}$ but subsequently fragments into three-bodies (D$^{+}$/D$^{+}$/O) via a sequential break-up mechanism. A corresponding feature was observed in the H$^{+}$/H$^{+}$/O yield following single-photon double ionization \cite{reedy2018dissociation} and has also been observed in SFI induced fragmentation of triatomic molecules such as OCS \cite{Rajput2018}.

\section{Signatures of trication formation in 40-fs pulse data}\label{sec:trication_formation}

The comparison of the $\beta$-KER plot for the cases of short (10~fs) and long (40~fs) pulse durations (see Fig.~\ref{fig:compare40fsVS10fs}) provided experimental support for the role of nuclear motion occurring during double ionization process. These data highlighted two significant differences. The first is a shifting of the feature observed at $\sim$7.5~eV to higher $\beta$ angles. This, as discussed in the main text, can be attributed to unbending dynamics that result in the water dication undergoing fragmentation from geometries close to linear. A second, higher KER, feature is discernible in the 40~fs data but is notably absent for the case of shorter pulses.

\begin{center}
	\begin{figure}
		\centering
	\includegraphics[width=\linewidth]{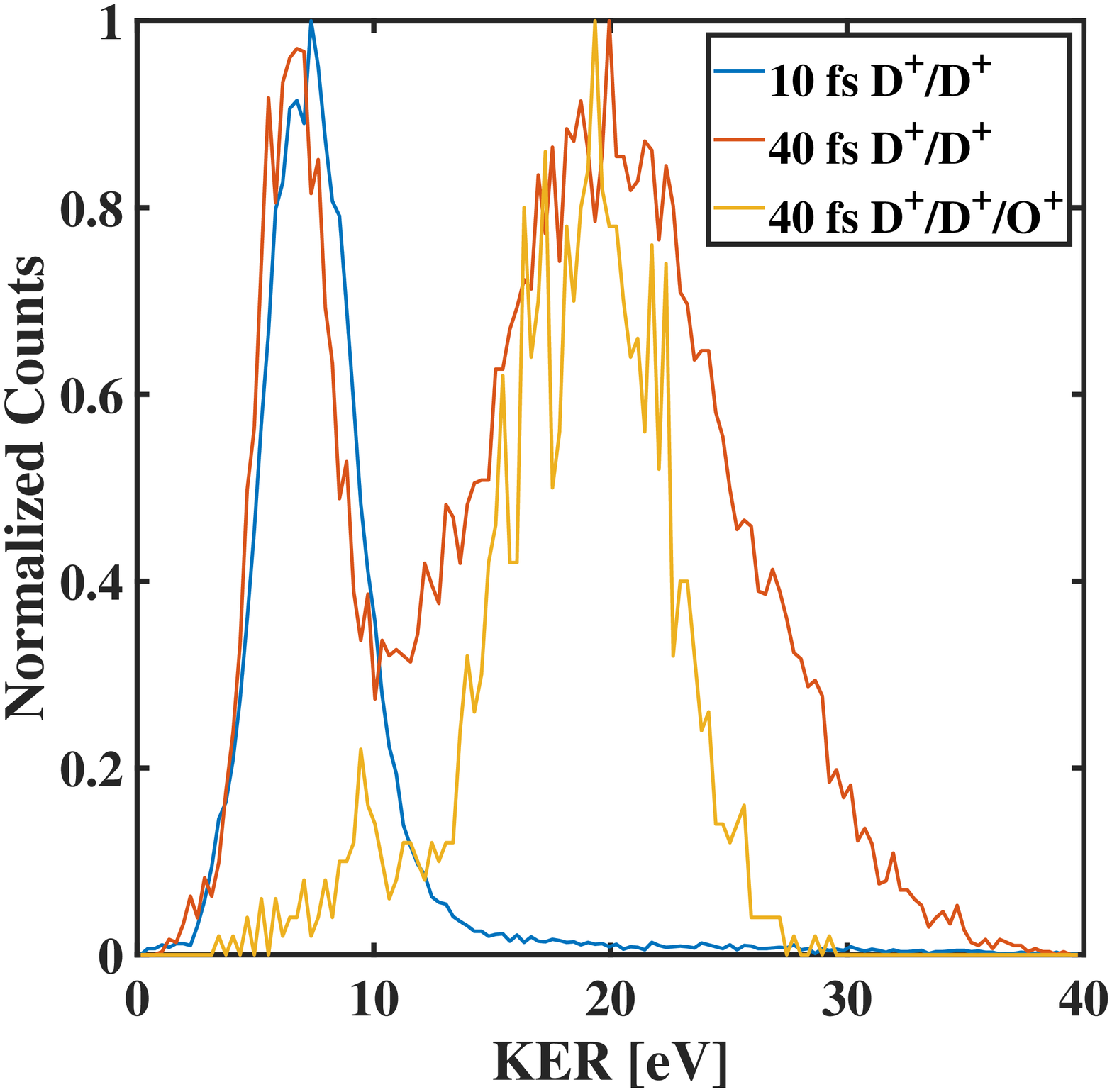}
		\caption{Kinetic Energy Release (KER) distribution of the D$^{+}$/D$^{+}$ dissociation channel for 10 (blue) and 40~fs pulse durations (red) and D$^{+}$/D$^{+}$/O$^{+}$ only for 40~fs pulse (yellow). The two D$^{+}$/D$^{+}$ lineouts correspond to $\beta$-angle-integrated versions of the data shown in Fig.\ref{fig:compare40fsVS10fs}. The overlap between D$^{+}$/D$^{+}$/O$^{+}$ channel and the D$^{+}$/D$^{+}$ in the 40~fs data indicates the high KER region are coming from events missing an O$^{+}$.}
		\label{fig:appd_10vs40KER}
	\end{figure}
\end{center}

In Fig.~\ref{fig:appd_10vs40KER} we present $\beta$-angle-integrated KER spectra for the two pulse durations. A striking difference is observed between the spectra, with the majority of the 40~fs counts originating from the feature centred at $\sim$20~eV. Given the significantly higher KER, a likely origin of this is fragmentation occurring from a more highly charged water ion and a missing coincidence partner ion (due to the limited detection efficiency of our apparatus). In order to verify this, we examined the D$^{+}$/D$^{+}$/O$^{+}$ coincidence channel and can unambiguously assign the origin of the higher KER feature to such a process.  

\bibliography{nonFCbib}

\end{document}